\documentclass[conference,10pt,letterpaper]{IEEEtran}
\usepackage[utf8]{inputenc} 
\usepackage[T1]{fontenc}
\usepackage{ifthen}
\usepackage{balance}
\usepackage{cite}
\usepackage[cmex10]{amsmath} 
\usepackage{amsmath, amssymb, dsfont, mathrsfs,amsthm,paralist}
\usepackage{float}
\usepackage{mathtools}
\usepackage{graphicx}
\usepackage{bm}
\usepackage{color}

\allowdisplaybreaks
\usepackage{multirow}

\usepackage{tikz}
\usetikzlibrary{plotmarks}
\usepackage{pgfplots}
\usetikzlibrary{calc}
\usetikzlibrary{shapes,arrows}
\usetikzlibrary{decorations.markings}
\usetikzlibrary{positioning}

\newcommand{\cW}{\mathcal{W}}

\newcommand{\R}{\mathbb{R}}

\newtheorem{thm}{Theorem}

\newtheorem{coro}{Corollary}

\newtheorem{defn}{Definition}
\newtheorem{remark}{Remark}

\newcommand{\erf}{\text{erf}}

\newcommand{\eps}{\varepsilon}

 \newcommand{\sgn}{\operatorname{sgn}}

\let\originalleft\left
\let\originalright\right
\renewcommand{\left}{\mathopen{}\mathclose\bgroup\originalleft}
\renewcommand{\right}{\aftergroup\egroup\originalright}

\pgfplotsset{compat=1.16}
\IEEEoverridecommandlockouts

\begin{document}

\title{Effects of Quantization on the \\ Multiple-Round Secret-Key Capacity}
\author{%
   \IEEEauthorblockN{Onur G{\"u}nl{\"u}\IEEEauthorrefmark{1},
                     Ueli Maurer\IEEEauthorrefmark{2},
                     and Jo\~ao Ribeiro\IEEEauthorrefmark{3}}
  \IEEEauthorblockA{\IEEEauthorrefmark{1}%
                     Chair of Communications Engineering and Security, University of Siegen, Germany, onur.guenlue@uni-siegen.de}
  \IEEEauthorblockA{\IEEEauthorrefmark{2}%
                      Department of Computer Science, ETH Zurich, Switzerland, maurer@inf.ethz.ch}
   \IEEEauthorblockA{\IEEEauthorrefmark{3}%
                     Department of Computing, Imperial College London, United Kingdom, j.lourenco-ribeiro17@imperial.ac.uk}
   \thanks{**Alphabetical author order.}
}

\maketitle

\begin{abstract}
   We consider the strong secret key (SK) agreement problem for the satellite communication setting, where a satellite chooses a common binary phase shift keying modulated input for three statistically independent additive white Gaussian noise measurement channels whose outputs are observed by two legitimate transceivers (Alice and Bob) and an eavesdropper (Eve), respectively. Legitimate transceivers have access to an authenticated, noiseless, two-way, and public communication link, so they can exchange multiple rounds of public messages to agree on a SK hidden from Eve. Without loss of essential generality, the noise variances for Alice's and Bob's measurement channels are both fixed to a value $Q>1$, whereas the noise over Eve's measurement channel has a unit variance, so $Q$ represents a channel quality ratio. We show that when both legitimate transceivers apply a one-bit uniform quantizer to their noisy observations before SK agreement, the SK capacity decreases at least quadratically in $Q$.
\end{abstract}

\section{Introduction}
The problem of secret key (SK) agreement consists in legitimate parties that observe dependent random variables to reliably agree on a key that is hidden from an eavesdropper by using a public communication link. We consider the source model for SK agreement where two legitimate parties, called Alice and Bob, and an eavesdropper, called Eve, observe $n$ independent and identically distributed (i.i.d.) realizations of random variables distributed according to a fixed joint probability distribution \cite{Mau93,AhlswedeCsiz}. The SK capacity, defined as the supremum of all achievable SK rates, is given in \cite{AhlswedeCsiz} for one-way public communication between legitimate parties. General upper and lower bounds on the SK capacity for two-way and multi-round public communication are also given in \cite{Mau93,AhlswedeCsiz}. Early results on the SK capacity use a weak secrecy-leakage metric that measures the normalized amount of information leaked about the SK to Eve. In \cite{MaurerWeaktoStrong,MaurerStrongSKRate}, lower and upper bounds on the SK capacity with a weak secrecy constraint are shown to be valid also with a strong secrecy constraint that is not normalized by the blocklength. Furthermore, improved lower and/or upper bounds on the SK capacity for general probability distributions are proposed, e.g., in \cite{MaurerUnconditional,RennerSK,Amin2010, WatanabeQKD}. A necessary and sufficient condition for the SK capacity to be positive for general probability distributions with two-way and multi-round public communication is provided in \cite{OrlitskyWigderson} and a sufficient condition in terms of Chernoff information is given in \cite{AminOurSKMulti}. Extensions to multiple parties are discussed in \cite{bizimBenelux,CsNar,Nitinawarat,benimmultientityTIFS,chan2018optimality} and capacity regions for SK agreement with privacy and storage rate constraints are given in \cite{csiszarnarayan, IgnaTrans, Tyagi, LaiTrans, bizimMMMMTIFS}.

As a binary example that provides interesting insights, SK agreement with a helpful satellite that is a remote source \cite[p.~118]{CsiszarKornerbook2011}, \cite[p.~78]{Bergerbook} whose outputs are measured through independent binary symmetric channels (BSCs) is considered in \cite{UeliProtocol,Mau93}. The satellite setting with BSCs illustrates that two-way and multi-round public communication, unlike one-way public communication, allows to achieve a positive SK rate even when both Alice's and Bob's noisy observations of the satellite outputs have a \emph{lower} quality than Eve's noisy observations. The conditions for this result to hold are that Eve's measurement channel should not be noiseless and Alice's and Bob's measurement channels should have positive channel capacities; see also \cite[Section 1.4]{JoaoUeliSK} for precise definitions. 

To achieve a positive SK rate with two-way and multi-round public communication advantage distillation protocols are used, including the repetition protocol \cite{Mau93}, the parity check protocol \cite{UeliProtocol,GanderMaurerPCP}, and other protocols such as in \cite{MuramatsuAdv, Liu2003Adv, AminOurSKMulti}. Advantage distillation protocols aim to provide an information-theoretic advantage to the legitimate parties by selecting a subset of their observed symbols for which legitimate parties have an advantage over Eve. To focus on scenarios where advantage distillation is necessary to reliably agree on a SK, a metric called \emph{channel quality ratio} is defined in \cite{JoaoUeliSK} as the maximum of the ratio of the capacity of the Eve's BSC vs. the capacity of Alice's or Bob's BSC. For the satellite setting with BSCs, the SK capacity is shown to decrease quadratically in the channel quality ratio when it is sufficiently large, achieved by using the parity check protocol \cite{JoaoUeliSK}. Furthermore, extensions of the satellite setting with BSCs to channels with binary phase shift keying (BPSK) modulated inputs and additive white Gaussian noise (AWGN) components are considered in \cite{MaurerUnconditional,NWMU09}, the former of which proves a sufficient condition to achieve a positive SK rate and the latter proves that using soft information increases the achievable SK rate as compared to a BSC that can be obtained by applying a one-bit uniform quantization at Alice and Bob.

For statistically independent AWGN satellite measurement channels with BPSK modulated inputs, we define below a new metric $Q$ that represents a channel quality ratio, which is a ratio of signal-to-noise ratios rather than the ratio of channel capacities defined for BSCs in \cite{JoaoUeliSK}. For any sufficiently large $Q$, we prove that the SK capacity is bounded from above by a term that decreases quadratically in $Q$ when a one-bit uniform quantization is applied at legitimate parties.

\section{Problem Definition and Main Results}
Consider a SK agreement problem where Alice and Bob who observe correlated random variables want to agree on a SK by using multiple rounds of public communication without a storage rate constraint such that the SK is hidden from Eve who also observes a correlated random variable. To obtain these correlated random variables we consider the following hidden source model, which is a sensible variation of the satellite setting defined in \cite{Mau93}. Suppose a binary remote source (or satellite) publicly chooses a strictly positive number $w\in\mathbb{R}^+$ and puts out either the symbol $R=+w$ or $R=-w$, i.e., BPSK modulated symbols, each with probability $1/2$. Without loss of generality, the antipodal satellite output is transmitted to Alice, Bob, and Eve through statistically independent zero-mean additive Gaussian noise channels with variances, respectively, $\sigma^2_A=\sigma^2_B=\sigma^2$ and $\sigma^2_E$, and Alice, Bob, and Eve observe i.i.d. random variables $X^n$, $Y^n$, and $Z^n$, respectively, where $n$ is the blocklength. 

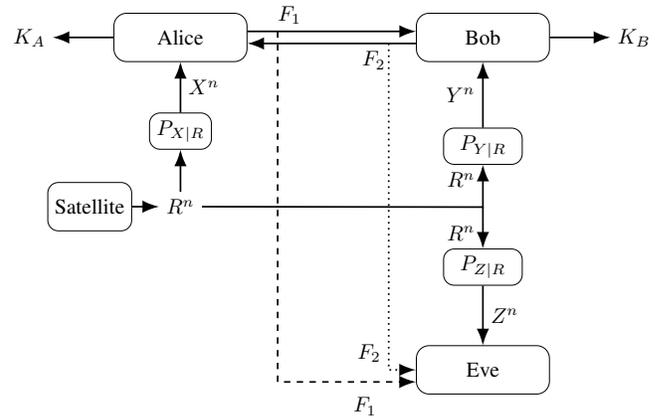
\begin{figure}
	\centering
	\resizebox{\linewidth}{!}{
		\begin{tikzpicture}
			\node (so) at (-1.5,-3.3) [draw,rounded corners = 5pt, minimum width=0.8cm,minimum height=0.8cm, align=left] {Satellite};
			\node (a) at (0,-0.5) [draw,rounded corners = 6pt, minimum width=2.2cm,minimum height=0.8cm, align=left] {Alice};
			\node (cY) at (5,-2.3) [draw,rounded corners = 5pt, minimum width=1.3cm,minimum height=0.6cm, align=left] {$P_{Y|R}$};
			\node (cZ) at (5,-4.3) [draw,rounded corners = 5pt, minimum width=1.3cm,minimum height=0.6cm, align=left] {$P_{Z|R}$};
			\node (f) at (0,-2.05) [draw,rounded corners = 5pt, minimum width=1cm,minimum height=0.6cm, align=left] {$P_{X|R}$};
			\node (b) at (5,-0.5) [draw,rounded corners = 6pt, minimum width=2.2cm,minimum height=0.8cm, align=left] {Bob};
			\node (g) at (5,-6) [draw,rounded corners = 6pt, minimum width=2.2cm,minimum height=0.8cm, align=left] {Eve};
			\draw[decoration={markings,mark=at position 1 with {\arrow[scale=1.5]{latex}}},
			postaction={decorate}, thick, shorten >=1.4pt] ($(a.east)+(0,0.1)$) -- ($(b.west)+(0,0.1)$) node [near start, above] {$F_1$};
			\draw[decoration={markings,mark=at position 1 with {\arrow[scale=1.5]{latex}}},
			postaction={decorate}, thick, shorten >=1.4pt] ($(b.west) - (0,0.1)$) -- ($(a.east)-(0,0.1)$) node [near start, below] {$F_2$};
			
			\node (a1) [below of = a, node distance = 2.8cm] {$R^n$};
			\draw[decoration={markings,mark=at position 1 with {\arrow[scale=1.5]{latex}}},
			postaction={decorate}, thick, shorten >=1.4pt] ($(cY.north)+(0.0,0)$) -- ($(b.south)+(0.0,0)$) node [midway, left] {$Y^n$};
			\draw[decoration={markings,mark=at position 1 with {\arrow[scale=1.5]{latex}}},
			postaction={decorate}, thick, shorten >=1.4pt] (so.east) -- (a1.west);
			\draw[decoration={markings,mark=at position 1 with {\arrow[scale=1.5]{latex}}},
			postaction={decorate}, thick, shorten >=1.4pt] (a1.north) -- (f.south);
			\draw[decoration={markings,mark=at position 1 with {\arrow[scale=1.5]{latex}}},
			postaction={decorate}, thick, shorten >=1.4pt] (f.north) -- (a.south) node [midway, right] {$X^n$};
			\draw[decoration={markings,mark=at position 1 with {\arrow[scale=1.5]{latex}}},
			postaction={decorate}, thick, shorten >=1.4pt] (a1.east) -- ($(cY.south)-(0,0.7)$) -- ($(cY.south)-(0,0.0)$) node [below left] {$R^n$};
			\draw[decoration={markings,mark=at position 1 with {\arrow[scale=1.5]{latex}}},
			postaction={decorate}, thick, shorten >=1.4pt]  ($(cY.south)-(0,0.7)$) -- ($(cZ.north)-(0,0.0)$) node [above left] {$R^n$};
			\draw[decoration={markings,mark=at position 1 with {\arrow[scale=1.5]{latex}}},
			postaction={decorate}, thick, shorten >=1.4pt] (cZ.south) -- (g.north) node [midway, right] {$Z^n$};
			\node (b2) [right of = b, node distance = 2.5cm] {$K_B$};
			\node (a2) [left of = a, node distance = 2.5cm] {$K_A$};
			\draw[decoration={markings,mark=at position 1 with {\arrow[scale=1.5]{latex}}},
			postaction={decorate}, thick, shorten >=1.4pt] (b.east) -- (b2.west);
			\draw[decoration={markings,mark=at position 1 with {\arrow[scale=1.5]{latex}}},
			postaction={decorate}, thick, shorten >=1.4pt] (a.west) -- (a2.east);
			\node (gwestmin) [below of = g, node distance = 0.2cm]{};
			\draw[decoration={markings,mark=at position 1 with {\arrow[scale=1.5]{latex}}},
			postaction={decorate}, thick, shorten >=1.4pt, dashed] ($(a.east)+(0.5,0.1)$) -- ($(a.east)+(0.5,-5.7)$) -- ($(a.east)+(0.5,-5.7)$) -- ($(g.west)-(0,0.2)$) node [below left=0.0cm and 1.5cm of gwestmin] {$F_1$};
			\draw[decoration={markings,mark=at position 1 with {\arrow[scale=1.5]{latex}}},
			postaction={decorate}, thick, shorten >=1.4pt, dotted] ($(b.west)-(0.45,0.1)$) -- ($(b.west)-(0.45,5.5)$) -- ($(b.west)-(0.45,5.5)$) -- (g.west) node [above left=0.0cm and 0.45cm of g.west] {$F_2$};
		\end{tikzpicture}
	}
\vspace*{-0.0cm}
	\caption{SK agreement with a helpful satellite for $\ell=1$ round of public communication.}\label{fig:ProblemDef}
	\vspace*{-0.0cm}
\end{figure}

Define the number of public communication rounds without a public-storage constraint as $\ell\geq 1$, which can be optimized for each parameter set. For $k=1,2,\ldots,\ell$, Alice creates public messages $F_{2k-1}$ according to some $P_{F_{2k-1}|X^n, F^{2k-2}}$, where $F^0$ is a constant, and sends the public messages to Bob. Similarly, Bob creates public messages $F_{2k}$ according to some $P_{F_{2k}|Y^n, F^{2k-1}}$ and sends the public messages to Alice. We remark that to create the random public messages a local source of randomness can be provided by using physical unclonable functions, which are unique and unclonable digital circuit outputs that are embodied by a device \cite{benimdissertation, bizimMDPI}, such as Alice's and Bob's encoders. After $\ell$ rounds of public communication, Alice generates a SK $K_A$ by using $(X^n, F^{2\ell})$ and Bob generates another SK $K_B$ by using $(Y^n, F^{2\ell})$. Alice and Bob aim to generate the same uniformly distributed key without leaking any information about it to Eve. Fig.~\ref{fig:ProblemDef} illustrates the SK agreement setting with a helpful satellite for $\ell=1$ round of public communication.

Without loss of generality, we focus on the setting where the Satellite-to-Eve channel noise component has a variance $\sigma^2_E=1$, whereas the Satellite-to-Alice and Satellite-to-Bob channel noise components both have variances $\sigma^2=Q>1$. Therefore, for this setting the legitimate receivers Alice and Bob observe $X^n$ and $Y^n$, respectively, that are \emph{lower} quality versions of the binary satellite outputs $R^n$ as compared to the quality of Eve's observations $Z^n$. This setting represents the general setting with having Alice, Bob, and Eve scale their observations by $\frac{1}{\sigma_E}$ such that $Q=\frac{\sigma^2}{\sigma^2_E}\geq 1$. Thus, $Q$ represents a \emph{channel quality ratio} between Alice or Bob's channel and Eve's channel, where the quality of Alice's and Bob's observations $X_i$ and $Y_i$, respectively, of the satellite output $R_i$ as compared to the quality of Eve's observation $Z_i$ degrades as $Q$ increases for all $i=1,2,\ldots, n$. 

We assume that all parties apply a one-bit uniform quantizer to its observed Gaussian symbols to obtain $X_q^n$, $Y_q^n$, and $Z_q^n$, where $X_q,Y_q,Z_q\in\{-w,+w\}$ since one-bit uniform quantizers result in $X_{q,i}=w\cdot\sgn(X_i)$, $Y_{q,i}=w\cdot\sgn(Y_i)$, and $Z_{q,i}=w\cdot\sgn(Z_i)$ for all $i=1,2,\ldots, n$. We next define the \emph{quantized SK capacity with a helpful satellite}, denoted as $S_\cW^q(Q)$.

\begin{defn}\label{def:SKcapc}
For the satellite setting depicted in Fig.~\ref{fig:ProblemDef} and for a fixed $w\in\mathcal{W}$ and $Q> 1$, a SK rate $S^q(w,Q)$ is achievable if, for any $\delta\!>\!0$, there are some $n\!\geq\!1$, encoders, decoders, and $\ell\geq 1$ for which $S^q(w,Q) = \log(|\mathcal{K}_A|)/n$ and
\begin{alignat}{2}
    &\Pr[K_A\neq K_B]\leq \delta &&\qquad\qquad (\text{reliability})\\
    &H(K_A)\geq nS^q(w,Q)-\delta &&\qquad\qquad (\text{uniformity})\\
    & I(K_A; F^{2\ell}, Z^n)\leq \delta &&\qquad\qquad (\text{strong secrecy}).
\end{alignat}
The quantized SK capacity with a helpful satellite is defined as
\begin{align}
     S^q_{\cW}(Q)=\sup_{w\in \cW} S^q(w,Q).\label{eq:analogSKcap}
\end{align}
\end{defn}

In practice, hardware implementations, e.g., of communication networks impose that any modulated symbol $R$ transmitted by the satellite can be chosen from a bounded set $\cW$, which can be large. Thus, we assume in the following that $\cW\subseteq \R^+$ is a bounded set.

We next provide an upper bound on the quantized SK capacity and for this result, we consider the case where the channel quality ratio $Q$ is large, which corresponds to the best case for Eve in terms of the respective observed symbol quality. We prove that the quantized SK capacity scales at most by $ O\left(\frac{1}{Q^2}\right)$ for every sufficiently large channel quality ratio $Q$.

\subsection{Main Results}
We next list the main results of this work. The proof of Theorem~\ref{theo:maintheo} is given in Section~\ref{sec:quantizedUpper}. The proof of Corollary~\ref{cor:maincorollary} below follows by combining a simple reduction argument with the result of Theorem~\ref{theo:maintheo}, which is explained below.

\begin{thm}\label{theo:maintheo} 
We have $\displaystyle S^q_\cW(Q)= O\left(\frac{1}{Q^2}\right)$ for every non-empty bounded set $\cW\subseteq\R^+$ and sufficiently large $Q>1$.
\end{thm}

\begin{remark}
    The bound on $S^q_\cW(Q)$ given in Theorem~\ref{theo:maintheo} directly implies that the same bound is valid also on $S^q(w,Q)$ for all $w>0$ since the bound on the $S^q_\cW(Q)$ follows for any non-empty bounded set $\cW\subseteq\R^+$.
\end{remark}

The assumption that Eve has to apply a one-bit uniform quantizer is not realistic as a passive attacker cannot be forced to apply a particular decoding method. Thus, we next remove the assumption that Eve has to apply any quantization to $Z^n$, whereas Alice and Bob still have to apply a one-bit uniform quantization to $X^n$ and $Y^n$, respectively. We show that the bound $ O\left(\frac{1}{Q^2}\right)$ on the quantized SK capacity with a helpful satellite $S^q_\cW(Q)$ is also a bound for the more realistic version with $X^n_q$, $Y_q^n$, and $Z^n$. 

\begin{coro}\label{cor:maincorollary}
For every non-empty bounded set $\cW\subseteq \R^+$ and sufficiently large $Q> 1$, the SK capacity for the case where Alice and Bob quantize but Eve does not quantize their corresponding observations can be upper bounded by $\displaystyle O\left(\frac{1}{Q^2}\right)$. 
\end{coro}

The proof of Corollary~\ref{cor:maincorollary} follows since allowing Eve to use more information than a one-bit quantizer output cannot increase the SK capacity. Thus, Corollary~\ref{cor:maincorollary} illustrates that the results of Theorem~\ref{theo:maintheo} follow also when we remove the assumption that Eve has to apply quantization.

\section{Quantized SK Capacity Upper Bound}\label{sec:quantizedUpper}
We consider the quantized SK capacity with a helpful satellite defined in Definition~\ref{def:SKcapc}, where Alice, Bob, and Eve apply a uniform one-bit quantization to each symbol of their noisy measurements $X^n$, $Y^n$, and $Z^n$, respectively, to obtain $X^n_q$, $Y^n_q$, and $Z^n_q$. The proof of Theorem~\ref{theo:maintheo} is provided below.

\begin{IEEEproof}[Proof of Theorem~\ref{theo:maintheo} ]
Applying a classic upper bound on the SK capacity from \cite[Theorem 2]{Mau93} \cite[pp. 6]{MaurerStrongSKRate}, we have
\begin{equation*}
    S^q(w,Q)\leq I(X_{q};Y_{q}|Z_q).
\end{equation*}
Therefore, it suffices to show that
\begin{equation*}
    \sup_{w\in\cW}I(X_{q};Y_{q}|Z_q)=O\left(\frac{1}{Q^2}\right)
\end{equation*}
when $Q\to\infty$. Assuming that $w$ is restricted to an arbitrary bounded set $\cW$, we can actually show that
\begin{equation}\label{eq:finW}
    S^q_\cW(Q)\leq \sup_{w\in\cW}I(X_{q};Y_{q}|Z_q)\leq \frac{1}{5Q^2}
\end{equation}
for any sufficiently large $Q$. To prove a weaker version of \eqref{eq:finW} for simplicity, we remark that $X_{q}$ and $Y_{q}$ correspond to noisy versions of a random bit $R$ measured through a BSC with crossover probability
\begin{equation}
    \eps=\frac{1}{2}\left(1-\erf\left(\frac{w}{\sqrt{2Q}}\right)\right)
\end{equation}
whereas $Z_q$ corresponds to a noisy version of the same random bit $R$ measured through another BSC with crossover probability
\begin{equation}
    \gamma=\frac{1}{2}\left(1-\erf\left(\frac{w}{\sqrt{2}}\right)\right)
\end{equation}
where $\displaystyle \erf(\cdot)$ is the error function defined as $\displaystyle \erf(z)=\frac{2}{\sqrt{\pi}}\int_0^z e^{-t^2}dt$. Define $H_b(p)=-p\log(p)-(1-p)\log(1-p)$ as the binary entropy function. Then, using this representation we have
\begin{align}
    &I(X_{q};Y_{q}|Z_q)=I(X_{q};Y_{q}|Z_q=w)\nonumber\\
    &\;=H(X_{q}|Z_q=w)-H(X_{q}|Y_{q},Z_q=w)\nonumber\\
    &\;=H_b(\eps\gamma+(1-\eps)(1-\gamma))-H(X_{q}|Y_{q},Z_q=w)\nonumber\\
    &\;=H_b\left(\eps\gamma+(1-\eps)(1-\gamma)\right)\nonumber\\
    &\;\;\quad-\left(\eps\gamma+(1-\eps)(1-\gamma)\right)\cdot H_b\left(\frac{\eps^2 \gamma+(1-\eps)^2(1-\gamma)}{\eps\gamma+(1-\eps)(1-\gamma)}\right)\nonumber\\
    &\;\;\quad-(\eps(1\!-\!\gamma)\!+\!(1\!-\!\eps)\gamma)\!\cdot\! H_b\left(\frac{\eps(1-\eps)}{\eps(1-\gamma)+(1-\eps)\gamma}\right)\label{eq:upperboundMIopen}.
\end{align}
Using (\ref{eq:upperboundMIopen}), for every $w>0$ we have 
\begin{align}
   & \lim_{Q\to\infty} Q^2 \cdot I(X_{q};Y_{q}|Z_q)=\frac{2w^4\left(1-\erf\left(\frac{w}{\sqrt{2}}\right)^2\right)^2}{\pi^2 \ln 2}\label{eq:mathematicaupper}
\end{align}
which can be obtained in a routine manner by combining the following series expansions
\begin{equation}
    \eps=\frac{1}{2}-\frac{w}{\sqrt{2\pi Q}}+O\left(\frac{1}{Q^{3/2}}\right)\label{eq:epsilonseries}
\end{equation}
as $Q\to \infty$,
\begin{multline}
    H_b(p)=\!1\!-\!\frac{2}{\ln 2}\left(p\!-\!\frac{1}{2}\right)^2 \!-\!\frac{4}{3\ln 2}\left(p\!-\!\frac{1}{2}\right)^4\\ \!+\!O\left(\left(p\!-\!\frac{1}{2}\right)^6\right)\label{eq:binaryentropyseries}
\end{multline}
which is expanded around $p=1/2$, and
\begin{align}
    &\left(\eps\gamma+(1-\eps)(1-\gamma)\right)=\left[\frac{1}{2}-(1-2\gamma)\left(\eps-\frac{1}{2}\right)\right],\label{eq:crossoverseries1}\\
    &\left(\frac{\eps(1-\eps)}{\eps(1-\gamma)+(1-\eps)\gamma}\right)\nonumber\\
    &\quad=\Bigg[\frac{1}{2}-(1-2\gamma)\left(\eps-\frac{1}{2}\right)-8\gamma(1-\gamma)\left(\eps-\frac{1}{2}\right)^2\nonumber\\
    &\qquad\qquad\qquad\qquad\qquad+O\left(\left(\eps-\frac{1}{2}\right)^3\right)\Bigg],\label{eq:crossoverseries2}\\
    &\left(\frac{\eps^2 \gamma +(1-\eps)^2(1-\gamma)}{\eps\gamma+(1-\eps)(1-\gamma)}\right)\nonumber\\
    &\quad=\Bigg[\frac{1}{2}-(1-2\gamma)\left(\eps-\frac{1}{2}\right)+8\gamma(1-\gamma)\left(\eps-\frac{1}{2}\right)^2\nonumber\\
    &\qquad\qquad\qquad\qquad\qquad+O\left(\left(\eps-\frac{1}{2}\right)^3\right)\Bigg]\label{eq:crossoverseries3}
\end{align}
which are expanded around $\eps=1/2$. Since $\cW$ is bounded, applying  (\ref{eq:epsilonseries})-(\ref{eq:crossoverseries3}) to (\ref{eq:upperboundMIopen}) yields (\ref{eq:mathematicaupper}) for all $w\in\cW$. Furthermore, using the following inequality \cite{boundonerf1,boundonerf2frombook}
\begin{align*}
    1-\erf(z)\leq e^{-z^2}
\end{align*}
we obtain 
\begin{align*}
    1-2e^{-z^2}\leq\erf(z)^2
\end{align*}
which gives the inequality
\begin{align}
    (1-\erf(z)^2)^2\leq 4e^{-2z^2}.\label{eq:boundonerfsquare}
\end{align}
Applying (\ref{eq:boundonerfsquare}) to the limit in (\ref{eq:mathematicaupper}) for $\displaystyle z=\frac{w}{\sqrt{2}}$, we obtain the upper bound
\begin{align}
    \lim_{Q\to\infty} Q^2 \cdot I(X_{q};Y_{q}|Z_q)\leq\frac{8w^4e^{-w^2}}{\pi^2 \ln 2}\label{eq:mathematicaupperafterineq}
\end{align}
for all $w\in\cW$, which is maximized at $w^*=\sqrt{2}$ with the value $\approx 0.6330$. Using this bound, the proof of Theorem~\ref{theo:maintheo} follows.
\end{IEEEproof}

\section{Conclusion}
We considered the strong SK agreement problem for the satellite communication setting with independent additive Gaussian noise components whose variances define a channel quality ratio. We proved that if the two legitimate parties apply a one-bit quantizer to their noisy observations, the quantized SK capacity decreases at least quadratically in the channel quality ratio.

We conjecture that the quantized SK capacity studied above and the \emph{unquantized} SK capacity, in which Alice and Bob do not apply any quantization to their inputs, behave differently. More precisely, we conjecture that the unquantized SK capacity decreases at most \emph{linearly} in the channel quality ratio, unlike the quantized SK capacity that decreases at least quadratically.

\section*{Acknowledgment}
O. G\"unl\"u was supported by the German Federal Ministry of Education and Research (BMBF) within the national initiative for ``Post Shannon Communication (NewCom)'' under the Grant 16KIS1004. Authors thank Chung Chan for pointing out, among other things, that the statement of Lemma~2 of the previous version of this paper explicitly assumes that the channel quality ratio is finite and the statement does not follow for the asymptotic case, unlike what was claimed in its proof, which makes the result not compatible with the outer bound that is given above for any sufficiently large $Q$. Authors also thank other reviewers who provided useful suggestions for the previous version of this paper.

\IEEEtriggeratref{29}
\bibliographystyle{IEEEtran}
\bibliography{ska-refs}

\end{document}